\documentclass[amssymb,amsmath,prd,twocolumn,superscriptaddress,nofootinbib]{revtex4-1}

\usepackage{graphicx}
\usepackage{bm}
\usepackage{amssymb,amsmath}
\usepackage{mathrsfs}
\usepackage{latexsym}
\usepackage{color}
\usepackage[normalem]{ulem}
\usepackage{dcolumn}
\usepackage[colorlinks=true,citecolor=blue,urlcolor=blue]{hyperref}
\usepackage[usenames,dvipsnames]{xcolor}
\usepackage{xspace}
\usepackage{graphicx}
\usepackage{multirow}
\usepackage{aas_macros}

\DeclareMathOperator{\BH}{BH}
\DeclareMathOperator{\NS}{NS}

\newcommand{\CIT}{\affiliation{Department of Physics, California Institute of Technology, Pasadena, California 91125, USA}}
\newcommand{\CITLab}{\affiliation{LIGO Laboratory, California Institute of Technology, Pasadena, California 91125, USA}}
\newcommand{\AEI}{\affiliation {Max Planck Institute for Gravitational Physics (Albert Einstein Institute), Am M\"uhlenberg 1, Potsdam 14476, Germany}}
\newcommand{\UP}{\affiliation {Institut f\"{u}r Physik und Astronomie, Universit\"{a}t Potsdam, Haus 28, Karl-Liebknecht-Str. 24/25, 14476, Potsdam, Germany}}

\begin{document}

\title{Using Equation of State Constraints to Classify Low-Mass Compact Binary Mergers}

\author{Jacob Golomb}
\email{jgolomb@caltech.edu}
\CIT \CITLab

\author{Isaac Legred}
\email{ilegred@caltech.edu}
\CIT \CITLab

\author{Katerina Chatziioannou}
\email{kchatziioannou@caltech.edu}
\CIT \CITLab

\author{Adrian Abac}
\email{adrian.abac@aei.mpg.de}
\AEI \UP

\author{Tim Dietrich}
\email{tim.dietrich@aei.mpg.de}
\UP \AEI

\date{\today}

\begin{abstract}

Compact objects observed via gravitational waves are classified as black holes or neutron stars primarily based on their inferred mass with respect to stellar evolution expectations. 
However, astrophysical expectations for the lowest mass range, $\lesssim 1.2 \,M_\odot$, are uncertain. 
If such low-mass compact objects exist, ground-based gravitational wave detectors may observe them in binary mergers. 
Lacking astrophysical expectations for classifying such observations, we go beyond the mass and explore the role of tidal effects. 
We evaluate how combined mass and tidal inference can inform whether each binary component is a black hole or a neutron star based on consistency with the supranuclear-density equation of state.  
Low-mass neutron stars experience a large tidal deformation; its observational identification (or lack thereof) can therefore aid in determining the nature of the binary components.
Using simulated data, we find that the presence of a sub-solar mass neutron star (black hole) can be established with odds $\sim 100:1$ when two  neutron stars (black holes) merge and emit gravitational waves at signal-to-noise ratio $\sim 20$. 
For the same systems, the absence of a black hole (neutron star) can be established with odds $\sim 10:1$.  For mixed neutron star-black hole binaries, we can establish that the system contains a neutron star with odds $\gtrsim 5:1$.  Establishing the presence of a black hole in mixed neutron star-black hole binaries is more challenging, except for the case of a $\lesssim 1\,M_{\odot}$ black hole with a $\gtrsim 1\,M_{\odot}$ neutron star companion.
On the other hand, classifying each individual binary component suffers from an inherent labeling ambiguity.
\end{abstract}

\maketitle

\section{Introduction}

Astronomical observations have revealed a diversity in compact objects with masses $\lesssim 3 \,M_\odot$.
Classifying these observations as black holes (BHs), neutron stars (NSs), or white dwarfs (WDs), requires identifying observational signatures that are unique to each type. 
For example, pulsars are identified as NSs~\cite{Gold:1969zh}, while unique electromagnetic spectrum or emission signatures can distinguish between NSs and BHs even if the mass is unknown, as is the case for accreting X-ray binaries~\cite{Remillard:2006fc, Eijnden:2021wop, Titarchuk:2023vfl}.
On the gravitational-wave (GW) side, classification is simplified by the fact that ground-based GW detectors are only sensitive to objects that do not disrupt or collide before reaching the detector sensitive band $\gtrsim 10\, \rm Hz$. 
For example, a pair of maximum compactness WDs each with mass $ 1.3 \, M_\odot$ and radius $1700\, \rm{km}$ collide at a GW frequency of $ \approx 1 \,\rm Hz$, see App.~\ref{app:disruption} for calculation details. 
However, even after excluding WDs, distinguishing between NSs and BHs is challenging because, unlike electromagnetic emission, their GW emission is more similar, as it is primarily determined by the object's mass. 

GW mass measurements in conjunction with astrophysical and nuclear physics can lead to preliminary classification indications. Causality limits NS masses $\lesssim 3\,M_\odot$~\cite{Rhoades:1974fn, Kalogera:1996ci}; more massive objects observed in GWs must be BHs.
Astronomical and nuclear constraints suggest that NSs do not reach this theoretical maximum, however.
Estimates of the maximum mass of stable nonrotating NSs~\cite{Tolman1939, Oppenheimer1939} range $2.0-2.5\,M_{\odot}$~\cite{Legred21, Rezzolla18, Dietrich20, Pang:2021jta, Raaijmakers:2021uju, Fonseca:2021wxt}; rigidly rotating NS can be $\sim 20\%$ more massive~\cite{Cook:1993qj}.
Based on these constraints, Refs.~\cite{LIGOScientific:2020aai,Essick:2020ghc,gw190814} argued that the GW190425~\cite{LIGOScientific:2020aai} primary was likely a NS, while the GW190814~\cite{gw190814} secondary was a BH.
However, it is unclear if stellar evolution creates NSs up to the maximum mass allowed by nuclear physics; little evidence for or against this scenario is observationally available~\citep{gwtc3pops}.

Switching to the full mass distribution, Galactic observations indicate that the observed NS population is strongly peaked at $\sim 1.4 \, M_{\odot}$, with a lower (upper) truncation near  $1.1  (2.0) \, M_{\odot}$~\cite{Alsing:2017bbc,2020RNAAS...4...65F}. The Galactic BNS population is narrower and peaked at $1.4\,M_{\odot}$~\cite{Alsing:2017bbc, Shahaf23}, though the impact of selection effects on these results is unclear. 
Neither result is consistent with the GW-observed NS mass distribution that displays no prominent peak at $1.4\,M_{\odot}$~\citep{Chatziioannou:2020msi,Landry:2021hvl,gwtc3pops}.
Electromagnetic observations suggest a scarcity or even absence of sub-$5 \, M_{\odot}$ BHs~\cite{Ozel:2010,Kreidberg:2012, Farr11,Shao:2022qws}, though candidates, subject to debate~\cite{vandenHeuvel:2020chh,Thompson:2020nbd,  El-Badry:2022}, exist~\cite{Thompson:2018ycv, Jayasinghe:2021}. 
The $2.6 \, M_{\odot}$ secondary in GW190814~\citep{gw190814} as well as galactic observations~\cite{Barr:2024wwl,Chen:2024joj} indicate that if a mass gap between NSs and BHs does exist, it is not empty~\citep{gwtc3pops}.
In the absence of unambiguous classification for $\sim 2-3\,M_{\odot}$ objects, Refs.~\cite{Fishbach20,gwtc3pops,Farah22} modeled the mass distribution of all sub-$10\,M_{\odot}$ objects and identified a feature at $\sim 2.4\, M_\odot$. Under the assumption of nonoverlapping NS and BH distributions, such a feature could signal the transition from the NS to the BH population.

In contrast to these astrophysics- and nuclear physics-informed considerations about the high end of the NS mass range, the low end remains uncharted. No widely-accepted \textit{astrophysical} process results in stellar remnants of either type with masses $\lesssim 1.2 \, M_\odot$~\citep{Suwa18, Lattimer01, Lattimer07}, although \textit{physically} cold NSs remain stable down to $\mathcal{O}(10^{-1})\,M_\odot$~\citep{Lattimer01, Lattimer07}.\footnote{The minimum mass of a hot proto-NS is however likely larger than that of a cold NS~\citep{Silva16, Lattimer07, Strobel01}.}
Radio and X-ray observations have led to NS candidates with masses $\sim 1.17\,M_{\odot}$~\cite{Martinez15} and $\sim 0.8\,M_{\odot}$~\citep{Doroshenko22}. Additionally, masses and eccentricities of \textit{Gaia} binaries suggest the existence of $\sim 1\,M_{\odot}$ NSs~\citep{Shahaf23}.
As for BHs, while sub-$1\,M_{\odot}$ BHs do not form through stellar collapse, early-universe density fluctuations and sufficiently dissipative dark matter could collapse into primordial BHs with masses in this range~\citep{Carr74, NovikovPBH1979}. Searches for subsolar mass compact objects with GWs have as of yet yielded no detections~\cite{LIGOScientific:2021job, LIGOScientific:2022hai,Nitz:2022ltl}. 
If such BHs do exist, they may be detectable by current and future GW detectors, and properties such as their masses and spins may be measurable~\citep{Wolfe23, Morr_s_2023}. 

Given these uncertainties, classification of potential sub-$1.2\,M_{\odot}$ GW candidates requires
an additional unique signature: matter effects.\footnote{On the electromagnetic side, matter effects manifest as counterparts, such as with GW170817~\cite{LIGOScientific:2017ync}, proving the presence of at least one NS and a 10:4 preference for two~\cite{Hinderer:2018pei, Coughlin:2019kqf, Coughlin:2018fis}. Absence of a counterpart does not necessarily rule out NSs, as detectability may be limited by beaming or prompt collapse~\cite{LIGOScientific:2020aai}.}  
GWs from mergers involving NSs carry the imprint of tidal interactions in the signal phase evolution~\citep{Chatziioannou20, Hinderer:2007mb, Flanagan:2007ix}. 
To leading order\footnote{Higher-order effects, such as dynamical tides~\cite{Hinderer:2016eia,Pratten:2021pro,Gamba:2022mgx}, also affect the waveform and can aid in distinguishing NSs and BHs.}, the effect is quantified by the dimensionless tidal deformability which depends on the nuclear equation of state (EoS) ($c = G = 1$):
\begin{equation}\label{eq:lambda}
    \Lambda \equiv \frac{2}{3}k_2 C^{-5}\,,
\end{equation}
 where $k_2$ is the quadrupole tidal love number, and $C=m/R$ is the compactness, the ratio of the NS mass $m$ to its radius $R$. Tidal interactions enter the GW phase to leading 5$th$ Post-Newtonian (PN) order~\citep{Flanagan:2007ix, Favata14} through $\tilde{\Lambda}$, a mass-weighted combination of the component tidal deformabilities. 
BHs in General Relativity have vanishing $k_2$, making $\Lambda $ a unique signature of the compact object nature~\citep{Taylor09, Chia21} \footnote{Beyond static tides and $\Lambda$, Kerr BHs have nonvanishing dynamical tides~\cite{Perry23}.}. Tidal information has previously suggested the presence of at least one NS in GW170817 based on disfavoring zero tides~\cite{LIGOScientific:2018hze}, EoS-independent relations~\cite{LIGOScientific:2019eut} and consistency of the tidal measurement with EoS inference~\cite{Essick:2019ldf}. 
Furthermore, Ref.~\citep{Chen20} showed that lack of tidal signature can be used to identify ${\sim 1-2\,M_{\odot}}$ BHs if they exist, though distinguishing between NSBHs and BBHs is more challenging if the BH has a higher mass~\cite{Brown22}.

\begin{figure}
    \centering
    \includegraphics[width=.49\textwidth]{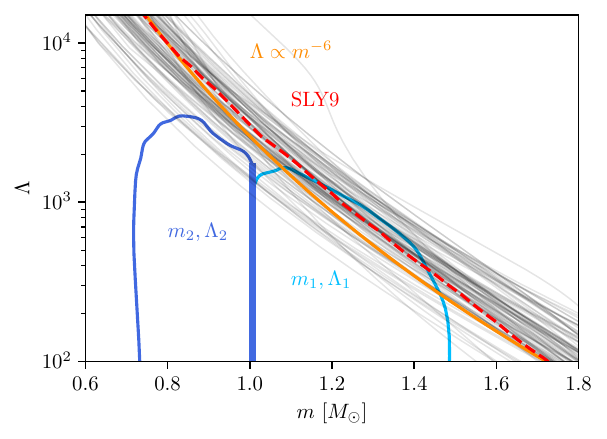}
    \caption{ The $m-\Lambda$ relation for draws from the EoS posterior from~\citep{Legred21} (gray lines). 
    A red dashed line denotes the SLY9 EoS. An orange solid line indicates the $\Lambda \propto m^{-6}$ trend. The posteriors of the masses and tidal deformabilties of the primary and secondary component of a BBH simulated signal are shown in light blue and dark blue, respectively. Despite poorer tidal constraints, the secondary is less consistent with the EoSs, suggestive of a BH. While this demonstration does not capture the full 4-dimensional mass-$\Lambda$ correlations, it sketches the main classification idea.}
    \label{fig:eos_macros}
\end{figure}

Tidal deformability becomes an increasingly better discriminator between BHs and NSs as the object's mass decreases. For $m\gtrsim 1 \, M_\odot$, $k_2$ scales as $k_2 \sim m^{-1}$~\citep{Zhao18}, resulting in $\Lambda(m) \sim m^{-6}$, see Fig.~\ref{fig:eos_macros}, assuming an approximately constant radius.\footnote{This is a good approximation excluding EoSs with phase transitions~\citep{Guillot13, Lattimer01}.} The lowest-mass NSs therefore exhibit the strongest tidal signatures and differ the most from BHs~\cite{Cullen:2017oaz}, with $\Lambda \sim \mathcal{O}(10^4)$ for $m\sim1\,M_{\odot}$, compared to $\Lambda \sim \mathcal{O}(10)$ for $m \gtrapprox 2 \,M_\odot$. 

In this work, we leverage the expected large tidal deformabilities of low-mass NSs, combined with astrophysically-informed EoS constraints to classify compact objects as either NSs or BHs. 
Our classification is based on the fact that a compact object's tidal deformability must be consistent with the EoS prediction if it is a NS, see the $m - \Lambda$ relation in Fig.~\ref{fig:eos_macros}, or zero if it is a BH.
While the true EoS is unknown, astronomical observations have placed constraints, giving independent predictions for the tidal deformability of a NS of a given mass, e.g.,~\cite{Dietrich20,Landry:2020vaw,Legred21,Pang:2021jta, Raaijmakers:2021uju}. 
This method expands upon efforts to identify NSs through a $\tilde \Lambda > 0$ condition~\cite{LIGOScientific:2018hze}, as we additionally require $\Lambda$ to be consistent with predictions from the dense-matter EoS, similar to the GW170817 classification of~\cite{Essick:2019ldf}. In other words, our analysis combines the discriminatory power of two conditions: BHs are consistent with $\Lambda=0$ and NSs are consistent with $\Lambda=\Lambda(m)$ as predicted by the EoS.

We test our classification approach with simulated data from low-mass sources with signal-to-noise ratios (SNRs) of 20 and 12 at advanced detector sensitivity. Lower (upper) limits on $\Lambda$ allow us to rule out a BH-BH (NS-NS) origin when at least one of the binary components is a NS (BH). Figure~\ref{fig:eos_macros} shows a demonstration of this idea in the BH-BH case. Though this plot is restricted to two dimensions and does not capture the strong correlations between $\Lambda_1$ and $\Lambda_2$, c.f., Fig.~\ref{fig:q1_comparisons}, the full-dimensional posterior structure is leveraged in the classification scheme laid out in Sec.~\ref{sec:eos-reweighting}.
In systems with sufficiently unequal masses, $m_2/m_1 \lesssim 0.8$, it might be possible to conclude that there is only a single NS. We also discuss an ambiguity in labeling individual objects that makes it difficult to identify the NS in a single-NS system.

The rest of the paper is organized as follows.
In Sec.~\ref{sec:methods}, we overview the parameter estimation methodology and source classification procedure. We present parameter estimation results on simulated signals in Sec.~\ref{sec:pe-tidal-prior}. Using these results, we quantify the evidence of BHs and NSs in Secs.~\ref{sec:has-bh} and~\ref{sec:NS-classification}, respectively. We conclude in Sec. \ref{sec:conclusions}.

\section{Methods}\label{sec:methods}

In this section, we describe the classification procedure and the methods for demonstrating its effectiveness. In Sec.~\ref{sec:pe}, we describe the simulated low-mass signals and parameter estimation. In Sec.~\ref{sec:eos-reweighting}, we fold in EoS information to quantify the probability of each source type.

\subsection{Classification-agnostic Parameter Estimation}\label{sec:pe}

We simulate data for binaries with all unique configurations of source-frame masses $(m_1, m_2)\in\{0.8, 0.9, 1.0, 1.1, 1.2\}\,M_\odot$ with $m_1 \ge m_2$  and source type NS-NS, BH-NS, NS-BH, and BH-BH, where the first (second) initial corresponds to the primary (secondary). 
The lower mass is selected both for computational reasons and because distinguishability is easier for even lower-mass systems.
This results in 55 total configurations.\footnote{The total number of possible systems is 100. Enforcing $m_1 > m_2$ and taking into account that equal-mass NS-BH and BH-NS systems are identical reduces this to 55.}
For brevity, we refer to BH-BH as BBH and NS-NS as BNS. 
We simulate sources with no spins and two network SNRs, one high-SNR set with $\rho_{\rm net} \approx 20$ and another lower-SNR set with $\rho_{\rm net} \approx 12$. 
The former corresponds to an optimistic detection scenario, although still quieter than GW170817~\citep{Abbott2017}, while the latter is representative of the bulk of detections. 
Further details are provided in App.~\ref{app:injection_properties}.
BHs are simulated with vanishing $\Lambda$. For NSs, we assign $\Lambda(m)$ according to their mass $m$ and the EoS SLY9~\citep{Gulminelli15}, chosen as a representative EoS that is consistent with current astronomical data~\citep{Legred21}, see Fig.~\ref{fig:eos_macros}.
We adopt standard priors for all parameters, detailed in App.~\ref{app:injection_properties}. 
We remain agnostic on source type and adopt a uniform prior between $0$ and $20 \times 10^{3}$ for the tidal deformabilities for all simulated signals.

We simulate data observed by the LIGO-Virgo detector network~\cite{observingscenarios, LIGOScientific:2014pky, VIRGO:2014yos} with a zero noise realization, corresponding to a geometric mean of many noise realizations~\cite{Nissanke:2009kt}. For the noise Power Spectral Densities (PSDs), we use the LIGO O4 low-sensitivity and O3 Virgo noise curves~\cite{observingscenarios,2020PhRvD.102f2003B,PhysRevLett.123.231108}. 
Signals are simulated and modeled with \textsc{IMRPhenomXAS\_NRTidalv3}~\cite{NRTidalv3}, a phenomenological, frequency-domain waveform model for the dominant GW emission from the coalescence of BNS mergers with aligned spin components. 
The model is based on a BBH GW model~\cite{Pratten:2020fqn}, which is then augmented with a closed-form tidal expression~\cite{Dietrich:2017aum,NRTidalv3}. 
The model incorporates dynamical tidal effects~\cite{Hinderer:2016eia} and is calibrated to a suite of numerical-relativity simulations. Two of these simulations are unequal-mass systems with a subsolar mass secondary ($0.98\,M_{\odot}$ and $0.90\,M_{\odot}$, with tidal deformabilities $\sim 2600$ and $\sim 4600$, respectively). The model has also been compared against an unequal-mass system with a subsolar mass component $\sim 0.94\,M_\odot$ and a tidal deformability of $\sim 9300$~\cite{Ujevic:2022qle}. Its reliability has been checked within $m_{1,2}\in[0.5,3.0]\,M_\odot$ and $\Lambda_{1,2}\in[0,20000]$, a range well-suited for our study.

\begin{figure}
    \centering
    \includegraphics[width=0.5\textwidth]{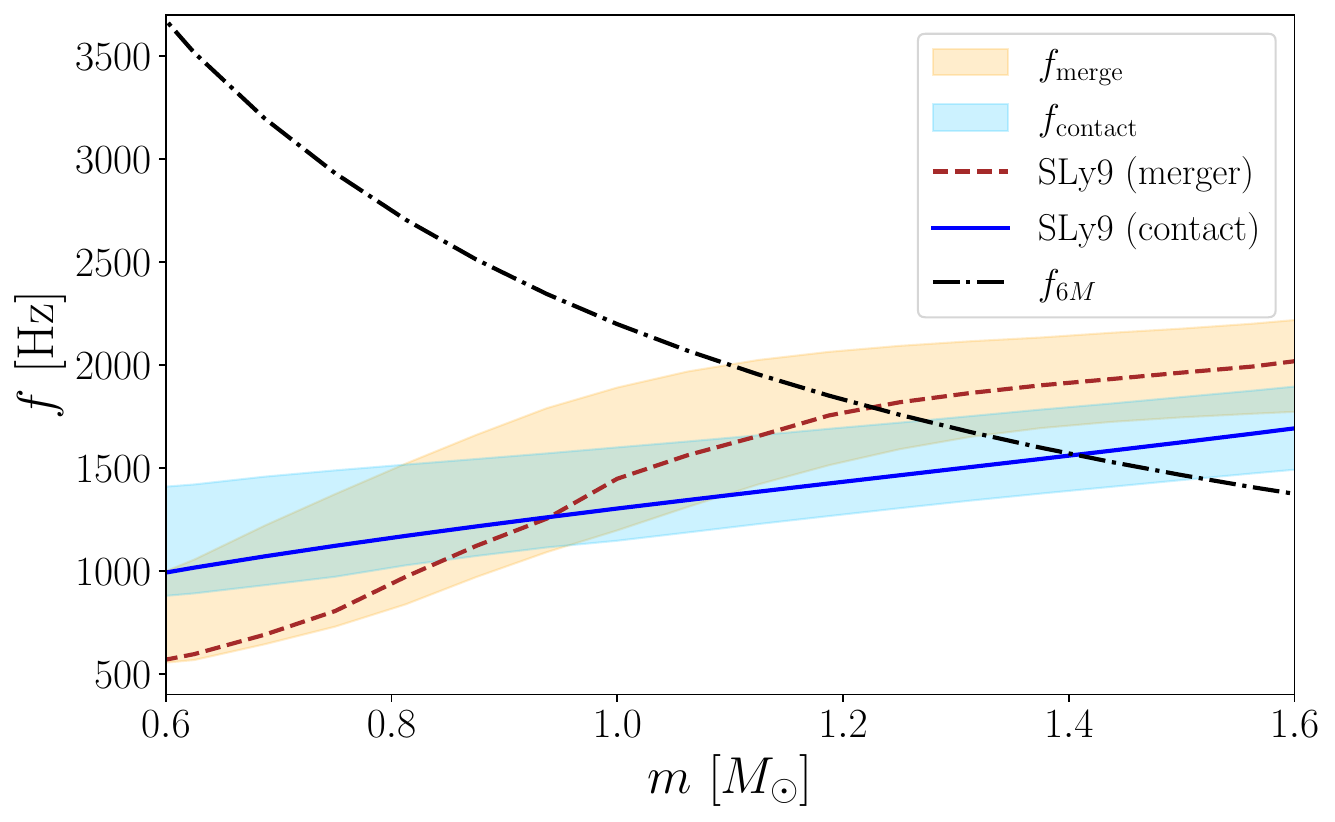}
    \caption{Relevant frequencies for late-inspiral signals: merger (peak strain, tan) and contact (orbital separation corresponding to objects touching, light blue) of NSs in equal-mass systems as a function of component mass.  Shaded regions correspond to marginalization over the EoS posterior from~\cite{Legred21}. Colored lines correspond to the SLy9 EoS~\cite{Douchin:2001sv, Gulminelli15}, which we use to simulate data. Lastly, we display an approximation for the plunge frequency of a comparable mass BBH $ f_{6M}$ with a black dash-dot line.  }
    \label{fig:bnsfrequencies}
\end{figure}

For illustrative purposes, we show relevant frequencies around the binary merger as a function of mass in Fig.~\ref{fig:bnsfrequencies}, see App.~\ref{app:disruption} for a detailed definition. 
We include the merger frequency, defined as the frequency of peak strain~\cite{Gonzalez:2022mgo}, the contact frequency, defined from a binary separation equal to the sum of the components' radii, and $ f_{6M} \equiv (6^{3/2}(m_1 + m_2))^{-1}/(2\pi)$, an approximation for the plunge frequency of BBHs.
In the mass range of interest, all frequencies are  between $\sim 1-3 \rm \, kHz$.

\subsection{Classifying Compact Binaries using EoS Information}\label{sec:eos-reweighting}

The possible source classes for each detected binary are $(T_1, T_2)$ one of $\rm \{(BH, BH), (NS, BH), (BH, NS), (NS, NS)\}$, where $T_1$ and $T_2$ refer to the source type (BH or NS) of the primary (more massive) or secondary (less massive) object, respectively.
For each event, the likelihood given an EoS $\epsilon$ and source type $T_1, T_2$ is obtained by marginalizing over the binary masses and tidal deformabilities:
\begin{equation}\label{eq:eos_event_posterior}
\begin{split} 
    \mathcal{L}(d|\epsilon, T_1, T_2) =& \int dm_1 \,dm_2 \,d\Lambda_1 \,d\Lambda_2 \,\mathcal{L}(d|m_1, m_2, \Lambda_1, \Lambda_2) \\&\times \pi(m_1, m_2) \pi(\Lambda_1, \Lambda_2| \epsilon, m_1, m_2, T_1, T_2)\,,
\end{split}
\end{equation}
where $\mathcal{L}(d|m_1, m_2, \Lambda_1, \Lambda_2)$ is the GW likelihood over the masses and tidal deformabilities, $\pi(m_1, m_2)$ is the prior on masses, and $\pi(\Lambda_1, \Lambda_2| \epsilon, m_1, m_2, T_1, T_2)$ is the prescription for computing the tidal deformabilities.
For EoSs with a single stable branch\footnote{If there are multiple stable branches we use a prior $\pi(\Lambda_i) = \sum_{j=0}^N\frac{1}{N}\delta (\Lambda_i- \Lambda(m_i|\epsilon, j))$, where $j$ indexes stable branches and $\Lambda(m_i|\epsilon, j)$ is the tidal deformability on the $j-$th branch. A NS of a given mass is equally likely to be formed on any stable branch.}
\begin{equation}
\pi(\Lambda_i|\epsilon, m_i, T_i) = \begin{cases} \delta(\Lambda_i - \Lambda(m_i|\epsilon))\,, & \text{if $T_i = $ NS} \\
\delta(\Lambda_i) \,, & \text{if $T_i = $ BH} \end{cases}\,.
\label{eq:Lambdaprior}
\end{equation}
Equation~\eqref{eq:Lambdaprior}  corresponds to the following prior on $\Lambda_i$: under the $T_i = \NS$ hypothesis, $\Lambda_i$ is determined by the EoS $\epsilon$ and $m_i$, whereas under the $T_i = \rm BH$ hypothesis, the object has a vanishing tidal deformability.
Equation~\eqref{eq:eos_event_posterior} is independent of the prior on $\Lambda_i$ and $m_i$ used in the original single-event analysis of Sec.~\ref{sec:pe} as it only depends on the single-event likelihood.
The $\Lambda_i$ prior in Eq.~\eqref{eq:eos_event_posterior} is instead the EoS-informed prior of Eq.~\eqref{eq:Lambdaprior}.

The mass prior is encoded in $\pi(m_1, m_2)$, which is selected to be uniform in the joint source-frame component mass space, with $m_1, m_2 \in [0.5, 1.8]\,M_{\odot}$.  
This uniform prior is chosen for simplicity, as no constraints exist on the mass distribution of $\leq 1.2 \,M_\odot$ NSs and BHs.
It is nonetheless consistent with constraints on the $\sim 1 - 2 \,M_\odot$ mass distribution~\cite{Landry:2021hvl, gwtc3pops}. 
If a population of low-mass binaries were discovered, the mass prior would also be inferred via an extension of Eq.~\eqref{eq:eos_event_posterior}, e.g,~\cite{Golomb22, Wysocki20}.

Whereas Eq.~\eqref{eq:eos_event_posterior} is conditioned on a single EoS $\epsilon$, the true EoS is unknown. We instead marginalize over the EoS and compute the likelihood for each classification:
\begin{equation}\label{eq:marginallikelihood}
P(d| T_1, T_2) = \int \mathcal{L}(d|\epsilon, T_1, T_2) \pi(\epsilon|d_{\rm aux}) d\epsilon \,,
\end{equation}
where $\pi(\epsilon|d_{\rm aux})$ is a distribution over EoSs informed by auxiliary data $d_{\rm aux}$.
We adopt the posterior from Ref.~\cite{Legred21} computed using a model-agnostic prior on the EoS based on a Gaussian process~\cite{Landry:2018prl, Essick:2019ldf,Legred22} and informed by radio-pulsar measurements~\cite{Fonseca:2021wxt, Antoniadis:2013pzd}, X-ray pulse-profile~\cite{J0030-Miller, J0030-Riley, J0740-Miller, J0740-Riley}, and GW observations~\cite{Abbott2017, LIGOScientific:2018hze, LIGOScientific:2020aai}. 
The EoS posterior is consistent with chiral effective field theory calculations at densities $\lesssim 1.5\, \rho_{\rm nuc}$ (where $\rho_{\rm nuc}$ is nuclear saturation density)~\cite{Weinberg:1978kz,Hebeler:2010jx, Tews:2018kmu, Drischler:2021kxf}, comparable to the central densities of $\sim 1$--$1.5\,M_{\odot}$ NSs, though it does not explicitly incorporate this information~\cite{Essick:2020flb}. It is also consistent with the existence of strong phase transitions~\cite{Essick:2023fso}.

The main physically relevant questions are
\begin{enumerate}
\item whether a source contains at least one BH,
\item whether a source contains at least one NS,
\item and, if so, whether it contains two NSs.
\end{enumerate}
Due to the lack of constraints on the merger rates of different source types in the relevant mass range we assign equal prior probability on 3 hypotheses ${\cal{H}}$: (i) the system has two NSs (BNS), (ii) the system has exactly one NS (OneNS), and (iii) the system has no NSs (BBH).

The marginal likelihood\footnote{The marginal likelihood is also commonly referred to as the ``evidence", though we use this term in its colloquial meaning.} of ${\cal{H}}$ is obtained by integrating over the relevant constituent source types:
\begin{equation}
    \mathcal{Z}_{{\cal{H}}} \equiv \int p(d|T_1, T_2) \pi(T_1, T_2|{\cal{H}})\, dT_1 \,dT_2\,,
\end{equation}
where $p(d|T_1, T_2)$ is given in Eq.~\eqref{eq:marginallikelihood}, and $\pi(T_1, T_2|H)$ is the normalized prior on the source types. 
The hypotheses ${\cal{H}} = \rm BNS$ and ${\cal{H}} = \rm BBH$ contain a single source type each, with the trivial priors $\pi(\NS, \NS|\rm BNS) = 1$, and $\pi(\BH, \BH|\rm BBH) = 1$ respectively.
The hypothesis ${\cal{H}} = \rm OneNS$ encompasses two source types, NSBH and BHNS, which we take to be equally likely \emph{a priori}, $\pi(\NS, \BH|\rm OneNS)  = \pi(\BH, \NS|\rm OneNS) =  1/2$. 

The marginal likelihood for whether the system contains at least one NS (``HasNS") is then
\begin{align}\label{eq:ZhasNS}
 \mathcal{Z}_{\rm HasNS} &= \mathcal{Z}_{\rm OneNS} \, \pi(\rm OneNS|\rm HasNS) \nonumber\\ &+\mathcal{Z}_{\rm BNS} \,\pi(\rm BNS|\rm HasNS)\,,
\end{align}
where $\pi(\rm OneNS|\rm HasNS) = \pi(\rm BNS| \rm HasNS) = 1/2$, meaning under the assumption the system has at least one NS, we assign an equal prior probability that it has one or two NSs. 
The marginal likelihood for whether the system contains at least on BH (``HasBH") is Eq.~\eqref{eq:ZhasNS}, with BNS $\rightarrow$ BBH.

In what follows, we present odds ratios between two hypotheses ${\cal{H}}_1$ and ${\cal{H}}_2$:
\begin{equation}\label{eq:oddsratio}
    \mathcal{O}^{{\cal{H}}_1}_{{\cal{H}}_2} = \frac{\mathcal{Z}_{{\cal{H}}_1}}{\mathcal{Z}_{{\cal{H}}_2}} \frac{\pi({\cal{H}}_1)}{\pi({\cal{H}}_2)}\,,
\end{equation}
where $\pi({\cal{H}})$ is the prior on the hypothesis ${\cal{H}}$, with $\pi(\rm HasNS)=\pi(\rm HasBH)=2\pi(\rm BNS)=2/3$.

\section{Measuring the Masses and Tides of Low-mass Compact Binaries}
\label{sec:pe-tidal-prior}

\begin{figure}
    \centering
    \includegraphics[scale=0.5]{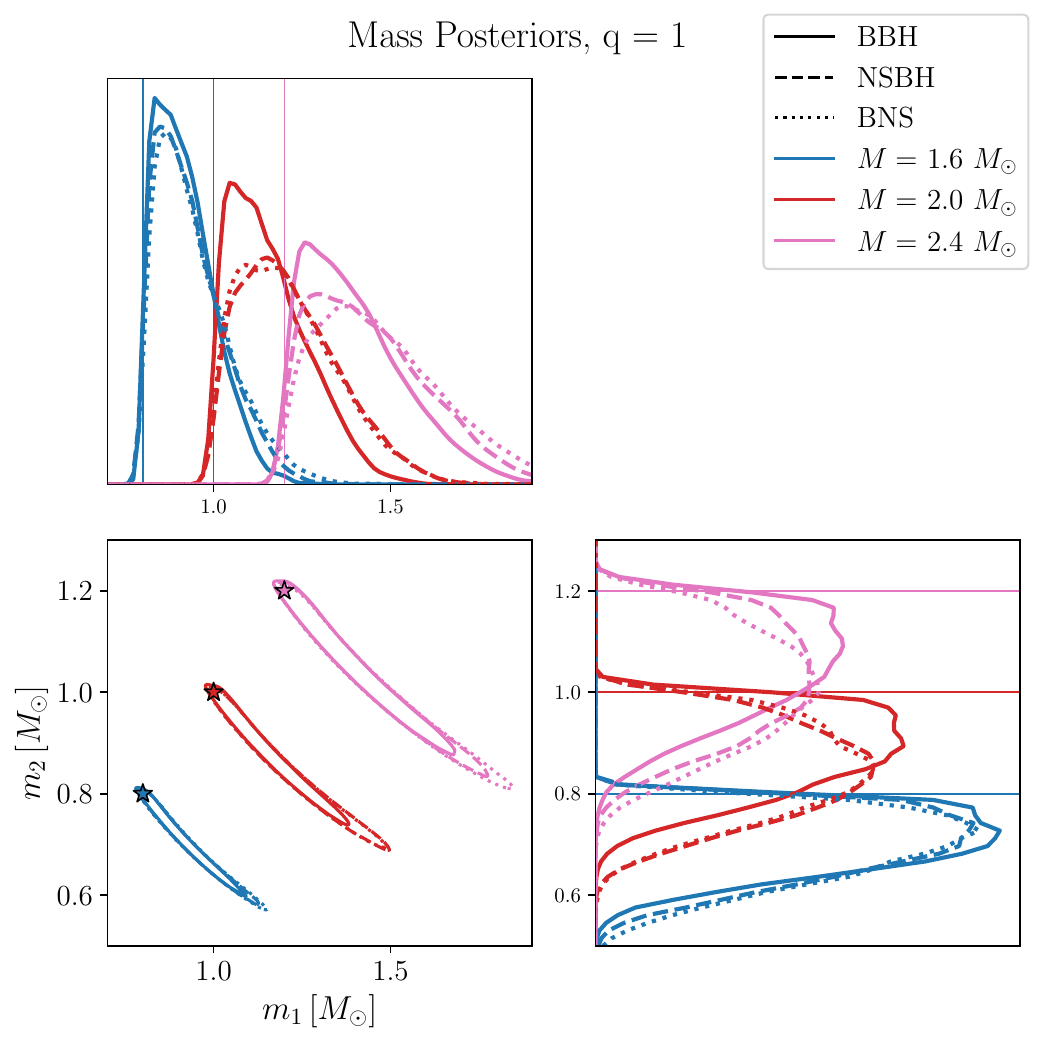}

    \caption{One- and two-dimensional marginalized source-frame mass posteriors for the $q \equiv m_2/m_1 = 1$ signals. Same-color lines denote systems with varying total mass $M$ with true values marked. For a given mass, varying line styles denote BBH, NSBH, and BNS systems. Contours represent two-dimensional 2-$\sigma$ regions. Given a simulated mass, similar posteriors across source types shows the subdominant effect of tides on the inferred masses. }
    \label{fig:q1_masses}
\end{figure}

In this section, we present posteriors from simulated signals. We do not assume we know whether each component is a NS or BH \textit{a priori}. Throughout, we present results from simulations with $\rho=20$.

The dominant intrinsic feature of a GW signal is the mass. 
In Fig.~\ref{fig:q1_masses}, we present marginal posteriors for the source-frame masses for select equal-mass systems. 
Measurement uncertainties are consistent with those of Ref.~\cite{Wolfe23}, c.f., their Figs.~1 and 2, at the same SNR.
Same-color lines denote systems with the same total mass, while varying line styles denote simulated source types. Same-mass signals result in similar mass posteriors, regardless of the source type,
with a minor trend for longer tails as the tidal effects increase.
This is due to the fact that the mass is primarily measured by the long inspiral phase (thousands of cycles), while tidal effect are relevant for the last $\sim20$ cycles. 
We obtain qualitatively similar posteriors for non-equal mass signals.

\begin{figure*}
    \centering
    \includegraphics[scale=0.35]{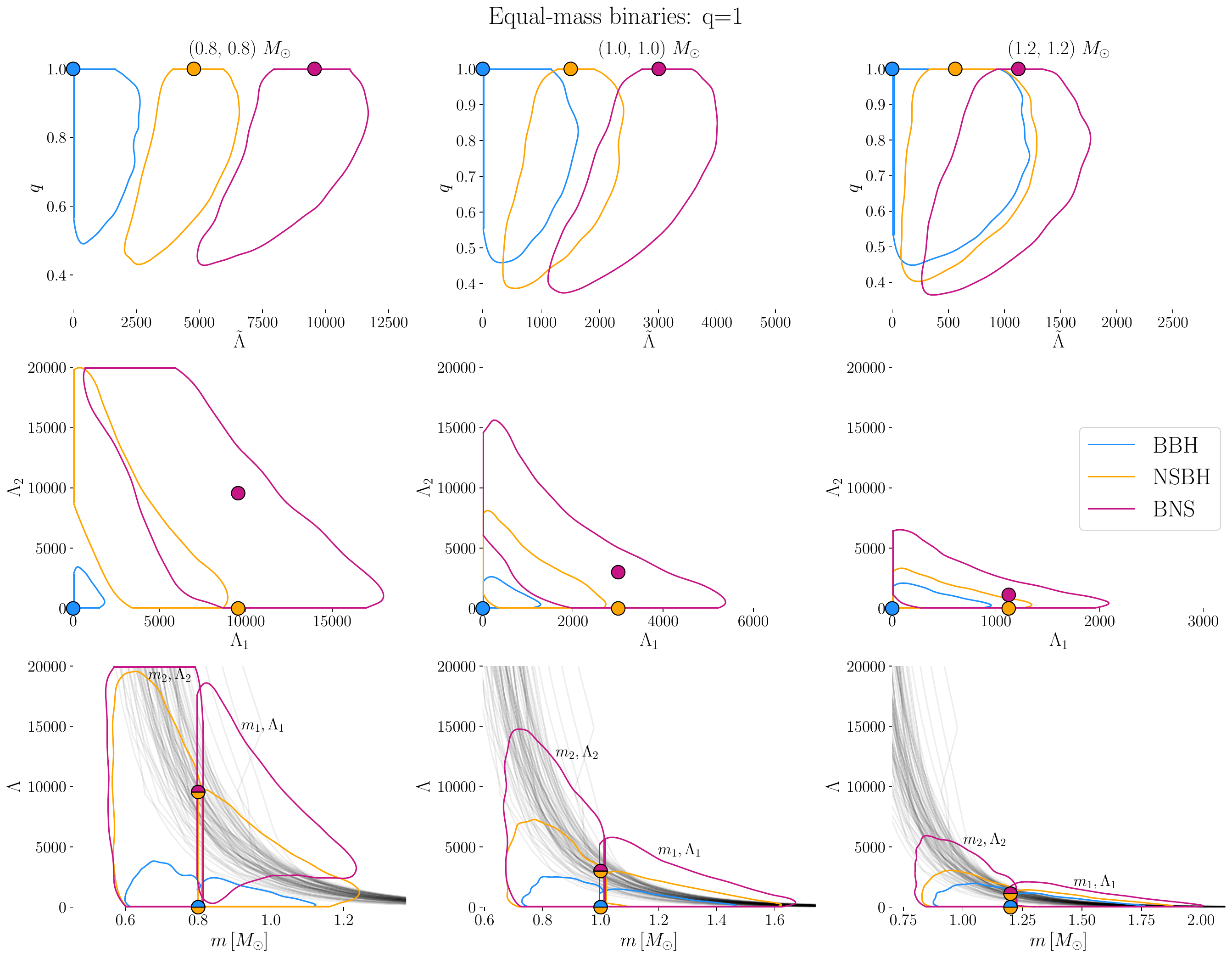}
    \caption{Two dimensional marginal posteriors for select parameters for systems with $q = 1$, with each column referring to a different simulated total mass. Blue, yellow, and magenta lines outline the 2-$\sigma$ contours of the posterior for the BBH, NSBH, and BNS systems, respectively. We omit the BHNS configuration as it is identical to NSBH for equal-mass simulations. The left (right) halves of the third row plots are the posterior of the primary (secondary), and include draws from the EoS distribution~\citep{Legred21} for reference. A decreasing total mass increases the tidal signature and correspondingly affects all posteriors.}
    \label{fig:q1_comparisons}
\end{figure*}

\begin{figure*}
    \centering
    \includegraphics[scale=0.35]{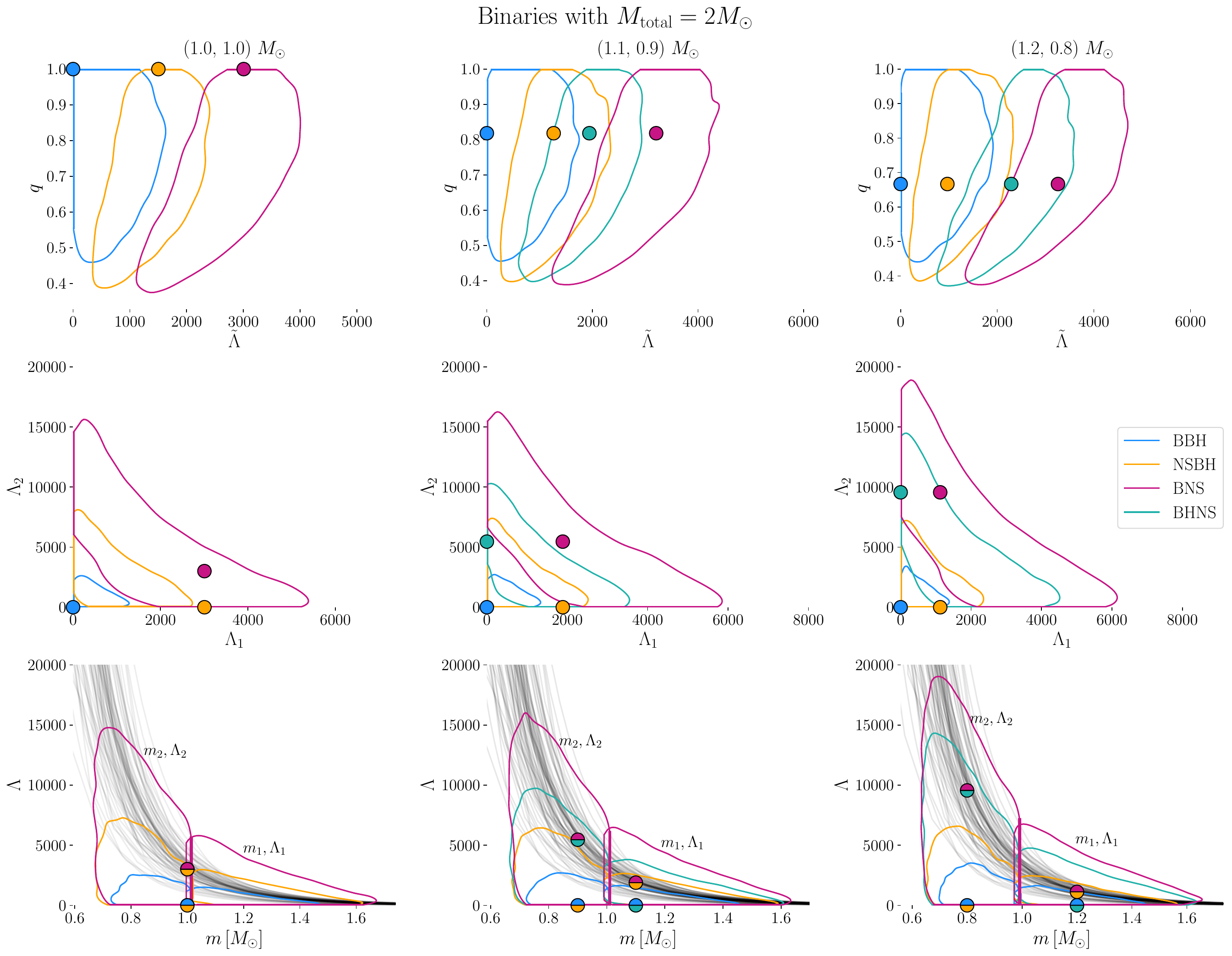}
    \caption{Similar to Fig.~\ref{fig:q1_comparisons} but for systems with the same simulated total mass $M = 2\,M_\odot$, with each column referring to a different simulated the mass ratio. When relevant, we also include BHNS configurations in green.  The posteriors of all parameters are, weakly sensitive to the true mass ratio, with the exception of the BHNS cases.}
    \label{fig:Mtotal2_comparisons}
\end{figure*}

Having established that the presence of tides does not strongly impact mass inference, we now turn to tidal inference.
Figures~\ref{fig:q1_comparisons} and~\ref{fig:Mtotal2_comparisons} show marginal posteriors for systems with fixed $q$ and $M$, respectively, with colors denoting the source type.
The top rows show the marginal $q - \tilde{\Lambda}$ posteriors.
All posteriors are consistent with the true (simulated) values. Within each panel, i.e., for configurations of the same mass, the posterior moves to higher values as the system contains more NSs and tidal effects become stronger.
The posteriors further show a positive correlation between $q$ and $\tilde{\Lambda}$ which becomes stronger as $\tilde{\Lambda}$ increases in value, consistent with~\cite{LIGOScientific:2018hze}.
An outcome of the increasing correlation strength is that the uncertainty also increases as the posterior is more extended both in the $q$, see also Fig.~\ref{fig:q1_masses}, and $\tilde{\Lambda}$ directions.

The $q-\tilde{\Lambda}$ posterior offers the first evidence about the presence/absence of tides and thus source classification. For all mass configurations, the BBH signals are consistent with the true value $\tilde{\Lambda} = 0$, and the posteriors are similar for different masses, c.f., blue contours in Figs.~\ref{fig:q1_comparisons} and~\ref{fig:Mtotal2_comparisons}, left to right. 
For NS-containing systems, the posteriors move away from $\tilde{\Lambda} = 0$, signaling the presence of tides. 
As expected, signals from lower-$M$ systems can rule out $\tilde{\Lambda} = 0$ with higher credibility due to their higher true $\tilde{\Lambda}$ value, c.f., yellow and magenta contours in Fig.~\ref{fig:q1_comparisons}, left to right. 
At a fixed $M$, the dependence of $\tilde{\Lambda}$ on the mass ratio is less pronounced, resulting in similar posteriors and thus ability to detect tides, c.f., yellow, green, and magenta contours in Fig.~\ref{fig:Mtotal2_comparisons}, left to right. 

Going beyond $\tilde{\Lambda}$, we turn to the tidal deformability of the individual binary components.
The second row of Figs.~\ref{fig:q1_comparisons} and~\ref{fig:Mtotal2_comparisons} shows posteriors for $\Lambda_1-\Lambda_2$.
The posteriors span much of the prior and show a strong anticorrelation consistent with~\cite{Chatziioannou:2018vzf,LIGOScientific:2018hze,gw170817eos}.
The direction of the anticorrelation is approximately a constant $\tilde{\Lambda}$ suggesting that almost all tidal information comes from measuring $\tilde{\Lambda}$, with limited higher-order information~\citep{Wade:2014vqa, Favata14}.
This is further demonstrated in App.~\ref{app:delta_lambda_tilde}. 
Consequently, $\Lambda_1-\Lambda_2 $ (second row) does not offer much additional information about the source type beyond $q-\tilde{\Lambda}$ (first row): exclusion of $\tilde{\Lambda}=0$ amounts to exclusion of $\Lambda_1=\Lambda_2=0$.
Crucially for source classification, all component tidal deformabilities are individually consistent with $\Lambda_i = 0$.\footnote{The only seeming exception is the lowest-mass BNS in Fig.~\ref{fig:q1_comparisons} but this is due to a posterior railing against the prior upper bound.}
Effectively, a $\tilde{\Lambda}$ measurement is ``spread" between $\Lambda_1$ and $\Lambda_2$ and the posterior for \emph{both} parameters is consistent with high values when \emph{either parameter} has a high true value. 

In the final row of Figs.~\ref{fig:q1_comparisons} and~\ref{fig:Mtotal2_comparisons}, we show the component $\Lambda_i - m_i$ posteriors, where gray lines are draws from the EoS posterior.
As expected from the second row, even in cases where $\tilde \Lambda=0$ is confidently ruled out, the posteriors are consistent with $\Lambda_i=0$.
More information can however be obtained by comparing the \emph{upper limit} on $\Lambda_i$ to EoS expectations at the relevant mass.
As expected, all BNS posteriors (magenta) are consistent with the EoS draws in both $(m_1, \Lambda_1)$ and $(m_2, \Lambda_2)$. 
Switching to the NSBH signals (yellow), the primary is always consistent with being a NS: for all masses nearly all the EoS draws fall within the yellow posteriors. In contrast and again for all mass configurations, about half the EoS draws fall within the posterior for the secondary binary component, indicating decreasing support for a NS interpretation.
Interestingly, this is despite the fact that the upper limit on $\Lambda_1$ is lower than that of $\Lambda_2$.
The expected tidal deformability increases so rapidly for lower masses that $\Lambda_1$ is more consistent with the EoSs than $\Lambda_2$.
The BHNS posteriors (green contours in Fig.~\ref{fig:Mtotal2_comparisons}) fully overlap with the EoS draws for all masses. This is because BHNSs have a larger $\tilde \Lambda$ than NSBHs for the same mass, pushing all upper limits to high enough values that are consistent with EoS predictions.

Finally, for BBH signals the posteriors for both components show some tension with EoS draws, which decreases with the total mass, c.f., blue contours of Fig.~\ref{fig:q1_comparisons}, left to right.
For the lowest mass configuration, c.f., left-most panel of Fig.~\ref{fig:q1_comparisons}, neither binary component overlaps with hardly any EoS draw.
In these cases, the GW data can constrain the tides to values that are too low compared to viable EoSs.
The binary mass ratio, on the other hand, does not strongly impact the overlap between the posterior and the EoSs, c.f., blue contours in Fig.~\ref{fig:Mtotal2_comparisons}, left to right. This is because the $\Lambda_i$ posterior does not strongly depend on the system mass, what changes is the EoS prediction which is a strong function of the total mass.

\section{Determining if a System Contains a Black Hole}\label{sec:has-bh}

\begin{figure*}
    \centering
    \includegraphics[width=0.8\textwidth]{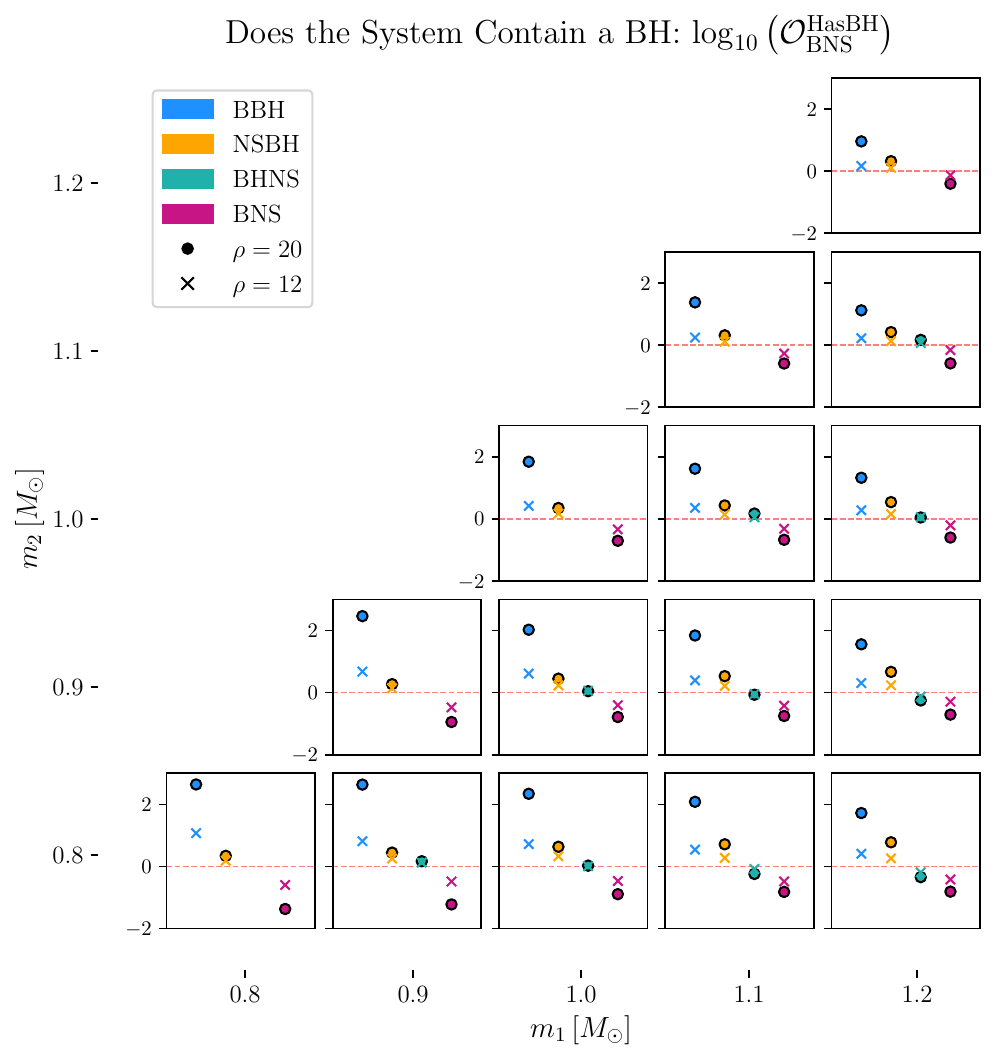}
    \caption{Base-10 logarithm of the odds ratio for each system containing at least one BH. Monte-Carlo errors for the odds ratios are too small to be visible in the scale of the figure. Panels correspond to the system source-frame masses and colors correspond to source type. The equal-mass panels do not contain BHNS systems as they are identical to the NSBH ones.  Dots (crosses) denote signals with SNR $20(12)$.
    Points above $\log_{10}(\mathcal{O^{\rm HasBH}_{\rm BNS}}) = 0$ (red dashed line) denote support for the presence of  at least one BH in the binary.
    }
    \label{fig:HasBH_odds}
\end{figure*}

Astronomical observations and nuclear physics considerations cannot directly motivate the nature of potential $\lesssim 1.2\,M_\odot$ GW detections such as the ones studied in Sec.~\ref{sec:pe-tidal-prior}. 
We undertake signal classification with the fundamental question: does
 the signal provide evidence for the \textit{presence} of a BH, thus establishing the existence of BHs below the expected astrophysical minimum mass?

We quantify this with the odds ratio $\mathcal{O}^{\rm HasBH}_{\rm BNS}$, where the ``HasBH" hypothesis consists of the BBH, NSBH, and BHNS source types with equal prior probabilities.
The alternative hypothesis is that the system is a BNS and thus the inferred 
 masses and tides of both objects must be consistent with the EoS.  In practice, the test comes down to whether the upper bound on the tidal effects is constraining enough to be in tension with the EoS prediction. 
We present the  base-10 logarithm  of the odds ratio, $\log_{10}\mathcal{O}^{\rm HasBH}_{\rm BNS}$ in Fig.~\ref{fig:HasBH_odds} for the $\rho=20$  (solid dots) and the $\rho=12$ signals (crosses). 
Below we focus on the $\rho=20$ results; we obtain qualitatively similar though weaker constraints when $\rho=12$.

The BBH signals (blue) show evidence for the presence of a BH, with odds $\gtrsim$10:1 for all masses. The evidence is stronger for lower-mass systems, with the odds ratio increasing from 10:1 to 100:1 between masses $1.2-1.2\,M_{\odot}$ and $0.8-0.8\,M_{\odot}$. 
This can be understood in the context of the EoS predictions; even though the $\Lambda$ posteriors are similar for all masses, c.f., blue contours in Fig.~\ref{fig:q1_comparisons}, bottom row, right to left, the EoS predicts that less massive NSs have much higher $\Lambda$ values. 
As the mass decreases, the
EoS predictions move away from the $(m, \Lambda)$ posterior support; this brings the data from less massive systems into more tension with the BNS hypothesis.

 NSBH signals (yellow) result in odds ranging between a few to $\sim 10:1$. 
For a given $m_2$, as $m_1$ increases (left to right), the odds ratio increases and we can more confidently infer the presence of a BH. This happens because both the true and inferred value of $\tilde \Lambda$ are smaller as $m_1$ increases.  
Both $\Lambda_1$ and $\Lambda_2$ are thus inferred to be smaller, but the estimate for $m_2$ is essentially unchanged, therefore, the secondary becomes more consistent with being a BH as the primary mass increases.
This contrasts with the case of increasing the total mass at constant mass ratio (bottom left to top right) where the inferred value of $\Lambda_2$ decreases and the inferred value of $m_2$ increases, so consistency with EoS predictions remains unchanged.  

Turning to the BHNS signals (green), we obtain near-equal odds for the presence of a BH for all masses. 
This is likely due to the larger tidal effects compared to the NSBH case (since now  the secondary is a NS) and the corresponding higher upper limits on tidal parameters, \emph{c.f.}, Fig.~\ref{fig:Mtotal2_comparisons}, allowing both objects to agree with the EoS predictions.  The odds for the presence of a BH decrease as the primary (BH) mass increases (left to right), as BHs and NSs become less distinguishable.

Finally, BNSs (magenta) always yield evidence against the presence of a BH, which decreases with the mass.

\section{Determining the Neutron Star Content of a System}\label{sec:NS-classification}

The complementary question is whether a system contains at least one NS and if yes, whether it contains two. 
Here, the evidence comes from both consistency of each object with EoS predictions and the exclusion of $\tilde{\Lambda}=0$.

\subsection{Does the System Contain a Neutron Star?}

\begin{figure*}
    \centering
    \includegraphics[scale=0.8]{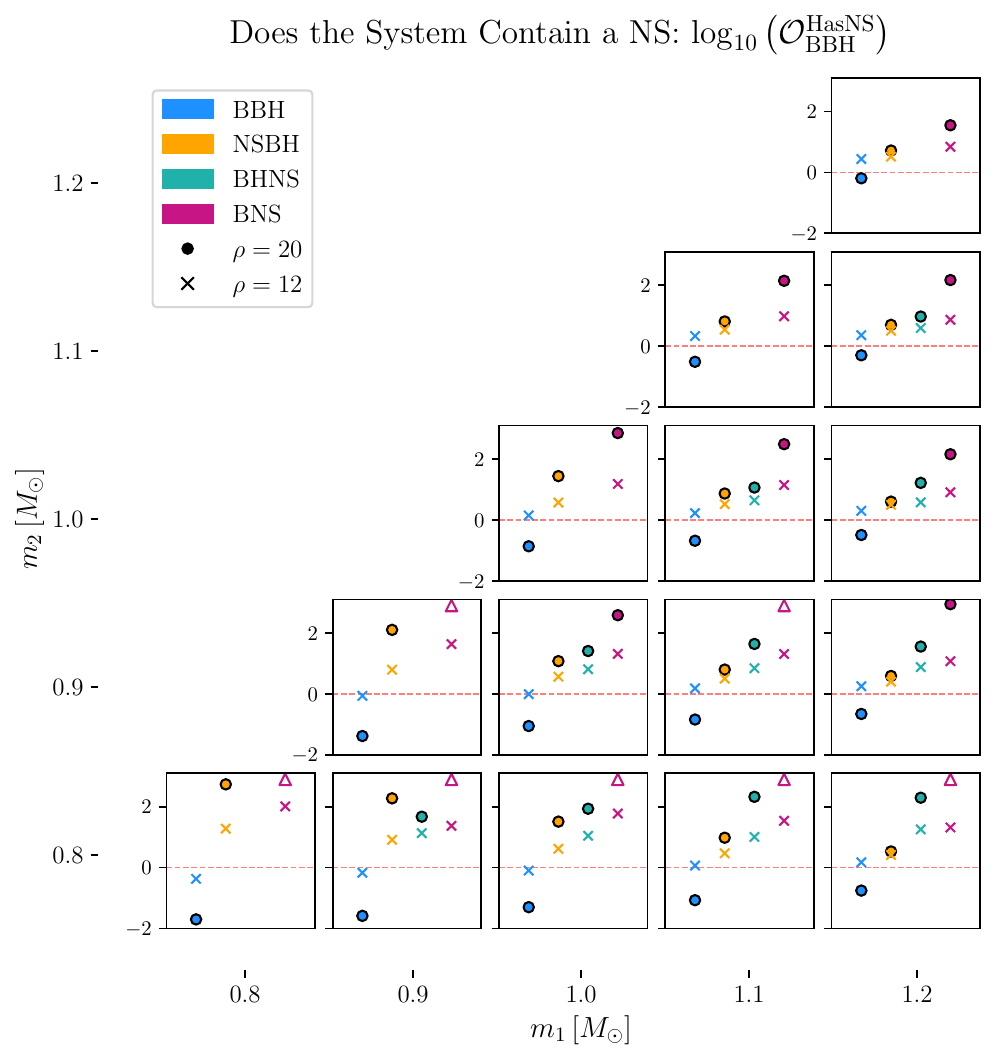}
    \caption{Similar to Fig.~\ref{fig:HasBH_odds} but for the odds ratio for each system containing at least one NS.  Points above $\log_{10}(\mathcal{O^{\rm HasNS}_{\rm BBH}}) = 0$ (red dashed line) denote support for the presence of at least one NS in the binary. Triangular markers indicate that the odds ratio lies somewhere above the y-axis limit.}
    \label{fig:HasNS_odds}
\end{figure*}

The evidence for whether there is at least one NS in a system is quantified with the odds ratio $\mathcal{O}^{\rm HasNS}_{\rm BBH}$, Eq.~\eqref{eq:oddsratio}. 
This is not equivalent to solely determining if the binary contains any matter; we further require the inferred tidal deformabilities to be consistent with the EoS.

In Fig.~\ref{fig:HasNS_odds}, we show $\log_{10}\mathcal{O}^{\rm HasNS}_{\rm BBH}$.
We again focus on the $\rho = 20$ results as $\rho = 12$ gives qualitatively similar, though less constraining, conclusions.
The log odds ratios for BBHs are negative, indicating that the data favor the absence of any NSs. 
As the mass decreases, so does the odds ratio 
from $\mathcal{O}^{\rm HasNS}_{\rm BBH} \approx 1/50$ for $1.2$--$1.2\,M_{\odot}$ to $\approx 2/3$  for $0.8$--$0.8\,M_{\odot}$. 
It becomes less plausible for the lowest-mass BBH systems to contain a NS as the signals lack the strong tidal signature that the EoSs predict for these masses, c.f., blue contours in Fig.~\ref{fig:q1_comparisons} bottom left compared to bottom right panel. 
All NS-containing systems yield $\log_{10}\mathcal{O}^{\rm HasNS}_{\rm BBH} > 0$ though again the evidence decreases as the NS mass increases.
For example, the odds ratio for $m_1 = 1.2 \,M_\odot,\, m_2 =0.8 \,M_{\odot}$ is $\mathcal{O}^{\rm HasNS}_{\rm BBH} \approx 4$, much lower than the $m_1 = m_2 = 0.8 \,M_\odot$ case which has $\mathcal{O}^{\rm HasNS}_{\rm BBH} > 100$.
At all masses, there is more evidence for a NS in BHNSs than NSBHs.
This is because the predicted tidal deformability of the primary is smaller than for the secondary, and thus a NS primary is more indistinguishable from a BH than a NS secondary. 
For systems containing exactly one  $\lesssim 1\,M_\odot$ NS, we obtain $\mathcal{O}^{\rm HasNS}_{\rm BBH} \gtrapprox 10$. 
The strongest evidence is obtained for the presence of a NS in the BNS systems, all of which have $\log_{10}\mathcal{O}^{\rm HasNS}_{\rm BBH} \gtrapprox 2$. 
This is consistent with the BNS posteriors of Fig.~\ref{fig:q1_comparisons} and~\ref{fig:Mtotal2_comparisons} that always rule out $\tilde{\Lambda} = 0$.

\subsection{Does the System Contain Two Neutron Stars?}

\begin{figure*}
    \centering
    \includegraphics[scale=0.8]{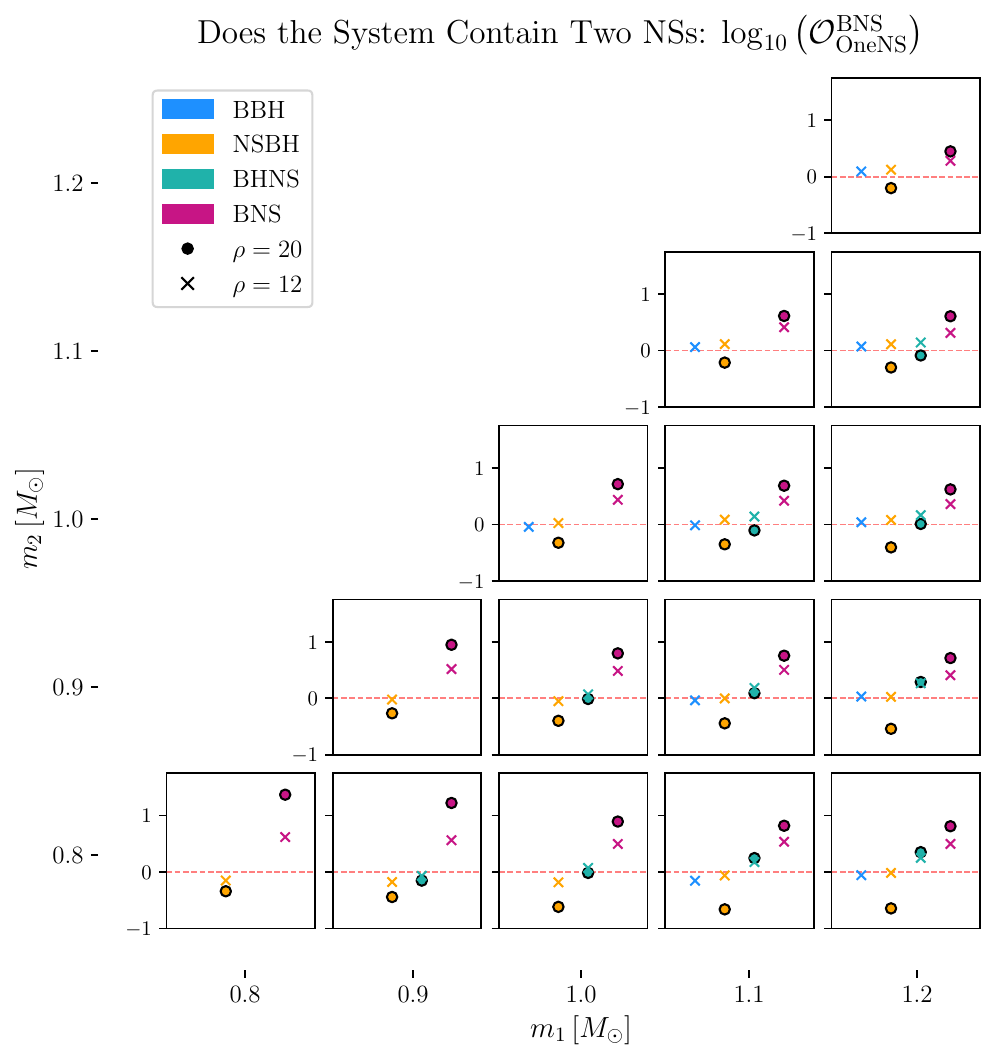}
    \caption{Similar to Fig.~\ref{fig:HasBH_odds} but for the odds ratio for each system containing exactly two NSs versus one NS. We only present results for systems with evidence of at least one NS in Fig.~\ref{fig:HasNS_odds} which includes all NS-containing systems. Points above $\log_{10}(\mathcal{O}^{\rm BNS}_{\rm OneNS}) = 0$ (red dashed line) correspond to systems that are more likely to have two NSs than one.
    }
    \label{fig:BNS_odds}
\end{figure*}

Having established the presence of a NS, the next question is whether the source is a BNS or it contains only one NS. 
We compare these two hypotheses with the odds ratio $\mathcal{O}^{\rm BNS}_{\rm OneNS}$.

We show results in Fig.~\ref{fig:BNS_odds}, restricting to systems with evidence for at least one NS in Fig.~\ref{fig:HasNS_odds} which in practice is all the NS-containing systems and a few BBHs with marginal evidence. We again focus on the $\rho=20$ results. BNS signals (pink) favor the presence of two NSs for all masses. 
As before, this evidence is stronger for less massive systems with odds $ \gtrapprox 10:1$ when both components are $\lesssim 1 \,M_\odot$.
NSBHs (yellow) provide stronger evidence against the presence of two NSs than BHNSs. 
This is again because determining the nature of the secondary (least massive) is easier than primary (most massive) component.

However, neither BHNS nor NSBH signals result in odds greater than 10:1 against the BNS hypothesis; the strongest evidence is obtained for the $1.2$--$0.8\,M_{\odot}$ NSBH binary with $\mathcal O^{\rm BNS}_{\rm OneNS} \sim 1/8 $.
The reason refers back to the posteriors in Figs.~\ref{fig:q1_comparisons} and~\ref{fig:Mtotal2_comparisons}. 
The BNS hypothesis requires that the EoS draws overlap with both the $(m_1, \Lambda_1)$ and $(m_2, \Lambda_2)$ posteriors. 
The bottom row of Figs.~\ref{fig:q1_comparisons} and ~\ref{fig:Mtotal2_comparisons} show that the EoS draws completely overlap the primary posterior for all NSBH (yellow) and BHNS (green) signals.
What is more, the posterior for the secondary is also fully (BHNS; green) or partially (NSBH; yellow) consistent with the EoS draws.

\subsection{If the System Contains One Neutron Star, is it the primary or the secondary?}

Though establishing the presence of exactly one NS is challenging at current sensitivity, we look forward to higher-SNR signals and consider how to identify which binary component it is.
Most analyses label objects based on relative mass, e.g., primary and secondary, hence the most straightforward approach is to examine whether the primary is a NS or a BH:
\begin{equation}
    \mathcal O^{\rm NSBH}_{\rm BHNS} =\frac{\mathcal Z_{\rm NSBH}}{\mathcal Z_{\rm BHNS}}\,.
\end{equation}
However, this suffers from a labeling ambiguity. 
For example, an equal-mass NSBH system is equally-well described by assigning the tides on either component. 
This is  due to the ambiguity in distinguishing binary components based on a property that is symmetric, i.e., the mass, and also plagues the component spins~\cite{Biscoveanu21}.

This ambiguity can be resolved by instead labeling the binary components with a unique property of each object that breaks this symmetry. 
For example, labeling binary components based on their tidal deformability would allow us to explore the properties of the stiffer and softer objects that reflect the NS and BH, respectively.
Such an approach is of course only applicable for systems with \emph{measurable} tidal asymmetry.  
For example, for BNSs, this approach would identify a ``stiff" and a ``soft" component, even if the tidal deformabilities are similar.
More generally, there is no guarantee that objects are in fact distinguishable, e.g., an equal-mass and nonspinning BBH, there is thus no generic strategy for extracting individual component properties.

\section{Conclusions}\label{sec:conclusions}

We have explored source classification for low-mass, $\leq 1.2\,M_\odot$ compact binary mergers based on the GW signal they emit and external information about the dense-matter EoS. 
The classification is based on the fact that the inferred component mass and tidal deformability must be consistent with EoS expectations if the object is a NS. 
A tidal measurement that is inconsistent with EoS predictions provides evidence that the object is not a NS, while $\Lambda = 0$ provides evidence for the object being a BH. 
The method's distinguishing power increases with decreasing mass, due to the fact that EoS predictions are a steep function of the mass, $\Lambda\sim m^{-6}$, and NSs become indistinguishable to BHs as the mass increases. 
Similarly, distinguishability is easier if the true EoS is stiffer as it would predict larger NS tidal deformabilities for all masses; here we have considered SLy9 that is consistent with the astrophysical data we employ.

We generally find it is easier to confirm the presence of a BH or NS than to refute it. 
For systems with subsolar-mass BHs, their presence can be identified at SNR $\rho=20$.  
In contrast, BNSs strongly disfavor the presence of a BH, with the evidence growing with decreasing masses. 
Complimentarily, signals from $\lesssim 1 \,M_\odot$ NS-containing binaries can reveal the NS presence based on compatibility of the mass-tidal measurement with EoS predictions.  
In contrast, if the binary \emph{does not} contain a NS, its presence is disfavored with the evidence again growing as the mass decreases.
Finally, identifying which object in a binary is a NS (or a BH) is subject to a labeling ambiguity that could be mitigated by labeling components based on relative tides rather than mass.
Higher-SNR signals due to detector upgrades~\citep{observingscenarios} or tighter EoS predictions thanks to future data will further strengthen distinguishability.

If subsolar-mass binaries exist and merge, combined mass and tidal information can aid in identifying the component nature and lead to constraints on primordial BH and NS physics. 
This prospect further motivates numerical simulations~\cite{Markin:2023fxx} and developing waveform models that can faithfully capture the large tidal effects of low-mass NSs.
It further motivates studies of alternative possibilities to BHs and standard NSs such as dark matter admixed NSs with lower tidal deformability~\citep{Hippert:2022snq}.
Tidal-based classification, as previously explored for higher-mass objects such as GW170817~\cite{LIGOScientific:2018hze,Essick:2019ldf,LIGOScientific:2019eut}, is especially promising for sub-solar mass objects whose nature is not otherwise astrophysically informed.
\\
\\

As this study was nearing completion, a preprint~\cite{Crescimbeni:2024cwh} that reached similar conclusions about the distinguishability of sub-solar mass BNS systems from BBHs appeared. Our methods differ in a few ways. The authors of~\cite{Crescimbeni:2024cwh} use Fisher matrix estimates (complemented with select full parameter estimation) and a modified \textsc{TaylorF2} approximant to account for NS disruption, as compared to our use of full parameter estimation (with priors that keep $\Lambda_1$ and $\Lambda_2$ positive) with the \textsc{NRTidalv3} waveform that includes appropriate termination conditions. 
Classification also differs: while Ref.~\cite{Crescimbeni:2024cwh} compares the upper limits on tidal inference to a fixed NS EoS, we form relevant hypotheses and marginalize over current uncertainty in the EoS to compute odds ratios. 
Additionally, we consider mixed NS-BH binaries, as opposed to only BNS and BBH systems. 
On the other hand, Ref.~\cite{Crescimbeni:2024cwh} also considers exotic compact objects.
Regardless, both studies find that we can tell apart a sub-solar mass BBH from a BNS at SNR $\gtrsim 12$.

\acknowledgments

We thank Kareem El-Badry for helpful discussions about astronomical observations of compact objects. We also thank Jocelyn Read for useful comments on the manuscript.
I.L. and K.C. acknowledge support from the Department of Energy under award number DE-SC0023101 and the Sloan Foundation. J.G. acknowledges funding from NSF Grant PHY-2207758.
The project was supported by the European Union (ERC, SMArt, 101076369). Views and opinions expressed are those of the authors only and do not necessarily reflect those of the European Union or the European Research Council. Neither the European Union nor the granting authority can be held responsible for them.
The authors are grateful for computational resources provided by the LIGO Laboratory and supported by National Science Foundation Grants PHY-0757058 and PHY-0823459.

Software: \texttt{bilby}~\cite{Ashton19, Romero-Shaw20}, \texttt{dynesty}~\cite{Dynesty}, \texttt{scipy}~\cite{scipy}, \texttt{numpy}~\cite{numpy}, \texttt{matplotlib}~\cite{matplotlib}, \texttt{lwp}~\cite{lwp}.

\appendix

\section{Limiting Frequencies}\label{app:disruption}

Compact binary inspirals terminate when the objects merge, disrupt each other, or their surfaces contact. In this Appendix, we quantify how compact binary components need to be in order to avoid disruption and contact and thus emit GWs in the sensitive band of ground-based detectors, see, e.g.,~\cite{Yamamoto23} for a similar calculation.

The onset of merger is not precisely defined, but a separation of $r=6M=6(m_1 + m_2)$ gives an order-of-magnitude estimate and a Keplerian frequency
\begin{equation}
    f_{\rm 6M} = \frac{1}{\pi}\sqrt{\frac{M}{(6M)^3}}\,,
\end{equation}
plotted in Fig.~\ref{fig:bnsfrequencies}; for $m_1 =m_2 = 1\,M_{\odot}$, $f_{\rm 6M}\sim2\,$kHz.  
Solar-mass compact objects therefore enter the LIGO-Virgo sensitive band before merger.

However, finite sizes might terminate the inspiral earlier if the objects contact each other  before reaching $r=6M$. 
For objects with radii $R_1$ and $R_2$, contact $r = R_1 + R_2$ occurs at a Keplerian frequency
\begin{equation}
    f_{\rm cont} = \sqrt{\frac{G (m_1 + m_2)}{4 \pi ^2 (R_1 + R_2)^3}}\,,
\end{equation}
also plotted in Fig.~\ref{fig:bnsfrequencies}.
For a BNS with $m_1=m_2=1\,M_{\odot}$ and $R_1 =R_2=12\,$km, $f_{\rm cont} \sim 1.5\,$kHz.
But for a NS-WD binary with an Earth-sized WD, $f_{\rm cont} \sim 0.2\, \rm Hz$, two orders of magnitude below the relevant frequency band.

Another possibility that prematurely ends an inspiral is disruption. The Newtonian tidal force felt by the secondary binary component due to the primary is
\begin{equation}\label{eq:newtonian_tides}
    F_{21} = \frac{G m_1 m_2 (r_{\rm out} - r_{\rm in}) (r_{\rm out} + r_{\rm in})}{(r_{\rm out} r_{\rm in})^2}\,,
\end{equation}
where $r_{\rm in} = r - R_2/2$ and $r_{\rm out} = r + R_2/2$ correspond to the distance between the primary and the outer and inner edge of the secondary, respectively. In the limit of wide orbital separation, $r \gg R_2$, Eq.~\eqref{eq:newtonian_tides} simplifies to
\begin{equation}
    F_{21} \approx \frac{2G m_1 m_2 R_2}{r^3}\,.
\end{equation}
The secondary disrupts when $F_{21}$ is comparable to its gravitational binding (self-)force
\begin{equation}
    F_{21} \approx \frac{Gm_2^2}{R_2^2} \,,
\end{equation}
which occurs at 
\begin{equation}\label{eq:disruption_radius}
    r \approx \left(2 \frac{m_1 R_2^3}{m_2}\right)^{1/3}\,,
\end{equation}
corresponding to a Keplerian orbital frequency of
\begin{equation}
    f_{\rm dis} \approx \sqrt{\frac{G m_2(m_1 + m_2)}{8 \pi^2 m_1 R_2^3}}\,.
\end{equation}

Therefore, 
\begin{equation}
    \left(\frac{f_{\rm dis}}{f_{\rm cont}} \right)^2 = \frac{m_2(R_1 + R_2)^3}{2m_1R_2^3}\,.
\end{equation}
For compact objects with comparable radii and masses, $f_{\rm dis}\approx 2 f_{\rm cont}$ and thus the binary contacts before disruption.
For a highly compact primary, for example a NS-WD binary with $R_1 \ll R_2$, $f_{\rm dis}< f_{\rm cont}$ and thus the binary disrupts before contact.
In any case, for binaries involving WDs, both of these frequencies are well below the LIGO sensitive band.

\section{Injection Properties}\label{app:injection_properties}

In this Appendix we provide more details for the parameter estimation analysis of Sec.~\ref{sec:pe}.
In Table~\ref{tab:injections} we list the extrinsic parameters of the simulated signals.  We select the luminosity distance unique to each system by scaling it to reach a target SNR, either 20 or 12. 
 
For the single-event analyses, we sample the parameter posterior using \textsc{Dynesty}~\citep{Dynesty} as implemented in \textsc{Bilby}~\citep{Ashton19, Romero-Shaw20}, with a prior that is uniform in component detector-frame masses and aligned spin components. 
We adopt standard isotropic priors for position and inclination parameters, and a luminosity distance prior that is uniform in  comoving volume~\cite{Romero-Shaw20}.
The prior on the component tidal deformabilities is uniform and ranges from $\Lambda = 0$ to $\Lambda = 20000$, the maximum value the waveform was validated on~\cite{NRTidalv3}. 
In some cases, the $\Lambda$ posterior distribution rails against this upper limit, but the simulated values for $\Lambda$ are always within in the prior bounds. 

We use a multibanding likelihood~\citep{Morisaki21} and analyze  512 or 256\,s of data (depending on the mass) at 8\,kHz with lower  and upper frequency cutoffs of 20\,Hz and 3.5\,kHz, respectively. The upper cutoff is above the inherent waveform termination~\cite{NRTidalv3, Gonzalez:2022mgo}.

\begin{table}
\begin{center}
\begin{tabular}{||c c c||} 
 \hline
 Parameter & Label & Value \\ [0.5ex] 
 \hline\hline
 Phase at 20\,Hz & $\phi$ & 0.24\,rad \\ 
 \hline
 Right Ascension & $\alpha$ & 0.18\,rad \\
 \hline
 Declination & $\delta$ & 0.62\,rad \\
 \hline
 Inclination & $\iota$ & 2.7\,rad  \\
 \hline
 Polarization Angle & $\psi$ & 0.58\,rad  \\
 \hline
 Merger time at geocenter & $t_c$ & 0\,sec (GPS) \\ [1ex] 
 \hline
\end{tabular}\label{tab:injections}
 \caption{Values for extrinsic parameters used for simulating the data.}
\end{center}
\end{table}

\section{Impact of measurements of $\delta\tilde{\Lambda}$}\label{app:delta_lambda_tilde}

\begin{figure*}[ht!]
    \centering
    \includegraphics[width=0.45\textwidth]{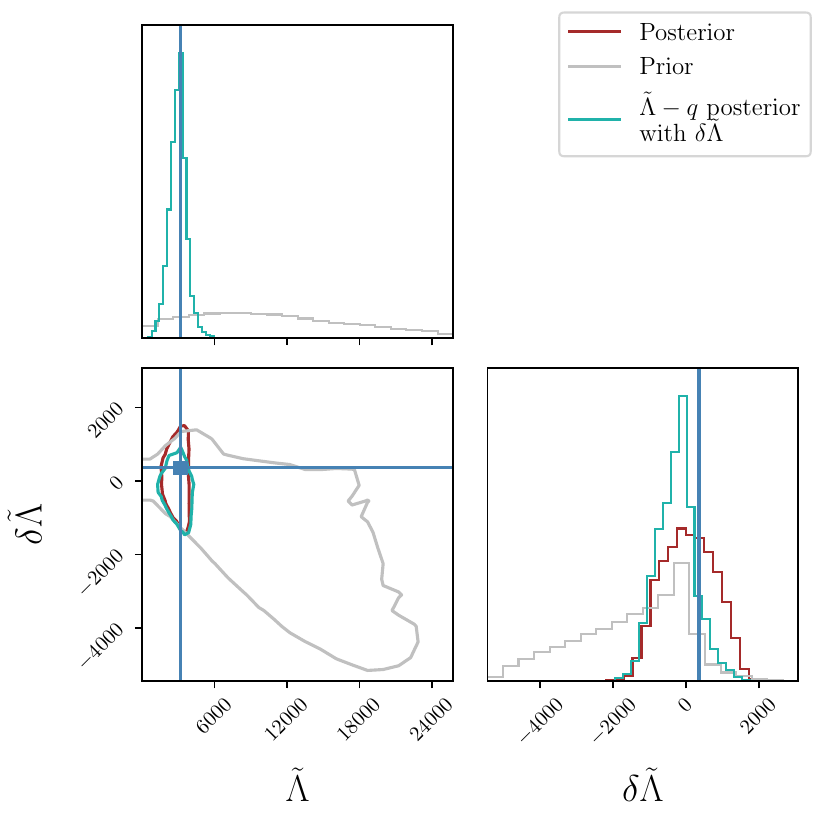}
    \includegraphics[width=0.45\textwidth]{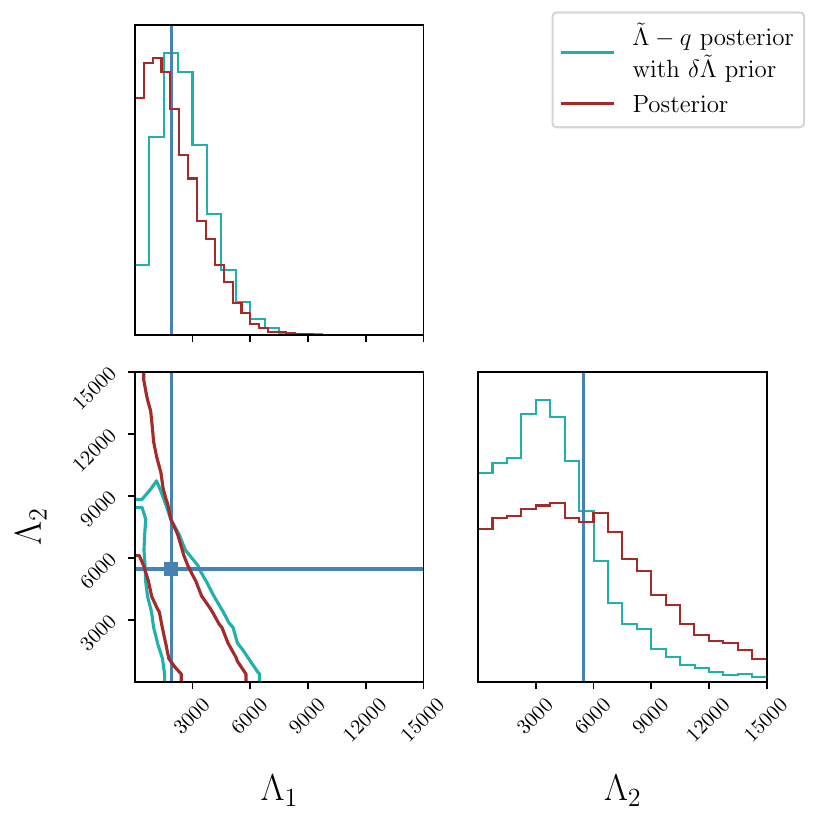}
    \caption{Marginal posterior (in brown) for tidal parameters from the BNS signal with $(m_1, m_2) = (1.1, 0.9)\,M_\odot$. (Left) Tidal parameters $\tilde{\Lambda}$ and $\delta \tilde{\Lambda}$, with the prior plotted in grey. (Right) Component tidal deformabilties $\Lambda_1$ and $\Lambda_2$. In both panels, the turquoise distribution corresponds to the posterior assuming that there is no information about $\delta \tilde{\Lambda}$.  We find that information about $\delta \tilde \Lambda$ is nonnegligible, though insufficient to break the degeneracy between $\Lambda_1$ and $\Lambda_2$.}
    \label{fig:lambdas_resampled}
\end{figure*}

In order to constrain the component tidal deformabilities, measurement of  an additional parameter beyond $\tilde \Lambda$, such as $\delta \tilde \Lambda$, is required.  The parameter $\delta \tilde \Lambda$ represents the tidal contributions to the frequency-domain phase which appear at 6PN and are not proportional to $\tilde \Lambda$; intuitively it is a measure of the asymmetry in the tidal contributions from the two components~\cite{Wade:2014vqa}.  
We examine the impact of the constraints on $\delta \tilde{\Lambda}$ in the tidal parameters from the $(m_1, m_2) = (1.1, 0.9) \,M_\odot$ BNS signal in Fig.~\ref{fig:lambdas_resampled}. 
In the left panel, we present the induced prior, see Sec.~\ref{sec:pe}, and the recovered marginal posterior for $\tilde{\Lambda}$ and $\delta \tilde{\Lambda}$. We obtain a symmetric 90\% credible interval for $\tilde{\Lambda} \in (1804, 4131)$ with respect to a prior that covers $0 < \tilde{\Lambda} \lesssim 26000$. 
In order to break the degeneracy between $\Lambda_1$ and $\Lambda_2$, we must measure additional parameters. However, $\delta \tilde{\Lambda}$ is relatively poorly measured at current sensitivity.   The left panel of Fig.~\ref{fig:lambdas_resampled}  shows that, even though the 1-d marginal posterior for $\delta \tilde \Lambda$ (red) appears to be well constrained relative to the prior (gray), this is primarily driven by $\tilde \Lambda $, c.f., the $2-d$ marginal posterior.   

In order to investigate how information about $\delta \tilde \Lambda$ impacts the component tidal deformabilities,
we approximate an inference where no information about $\delta\tilde \Lambda $ exists.  We draw  $(q, \tilde{\Lambda})$ samples from the full posterior and combine them with samples of $\delta \tilde{\Lambda}$ from its effective prior implied by the given $(q, \tilde{\Lambda})$, subject to the condition $\Lambda_i(q, \tilde \Lambda, \delta \tilde \Lambda) > 0$.  We display the marginal  distribution in the left panel panel of Fig.~\ref{fig:lambdas_resampled} (teal).
We compare this to the full marginal posterior on $\Lambda_1-\Lambda_2$ (red).
We find that while knowledge of $\delta \tilde \Lambda$ does change the distribution on $\Lambda_1$--$\Lambda_2$, this information does not substantially change the correlation structure. 
As expected for a well-measured parameter, this procedure leaves the $\tilde{\Lambda}$ posterior unaffected (left).
The measurement of $\delta \tilde{\Lambda}$ itself favors higher values of $\delta \tilde{\Lambda}$ (left), which correspond to  higher values of $\Lambda_2$ and lower values for $\Lambda_1$ (right). 

\bibliography{bibliography}
\end{document}